\documentstyle[12pt,preprint,aps,tighten,epsfig,rotate,floats]{revtex}

\def\beq{\begin{equation}}
\def\eeq{\end{equation}}
\def\beqa{\begin{eqnarray}}
\def\eeqa{\end{eqnarray}}
\def\be{\begin{equation}}
\def\ee{\end{equation}}
\def\bea{\begin{eqnarray}}
\def\eea{\end{eqnarray}}
\begin{document}

\preprint{
\vbox{\hbox{UWThPh-2001-31}}}
\title{Determination of $\langle (\pi\pi)_{I=2}|{\cal
Q}_{7,8}|K^0\rangle$ in the Chiral Limit}
\author{Vincenzo Cirigliano}
\address{Inst. f\"ur Theoretische Physik,
University of Vienna \\
Boltzmanngasse 5, Vienna A-1090 Austria \\
vincenzo@thp.univie.ac.at }
\author{John F. Donoghue and Eugene Golowich}
\address{Physics Department, University of Massachusetts\\
Amherst, MA 01003 USA \\
donoghue@physics.umass.edu,
golowich@physics.umass.edu \\}
\author{Kim Maltman}
\address{Department of Mathematics and Statistics, York University\\
4700 Keele St., Toronto ON M3J 1P3 Canada \\
and \\
CSSM, University of Adelaide \\
Adelaide, SA 5005 Australia \\
kmaltman@physics.adelaide.edu.au}
\maketitle
\thispagestyle{empty}
\setcounter{page}{0}
\begin{abstract}
\noindent
We reconsider the dispersive evaluation of the weak matrix
elements $\langle (\pi\pi)_{I=2}|{\cal Q}_{7,8}|K^0\rangle$
in the chiral limit. The perturbative matching is accomplished
fully within the scheme dependence used in the two loop
weak OPE calculations. The effects of dimension eight (and higher
dimension) operators are fully accounted for. We perform a numerical
determination of the weak matrix elements using our dispersive sum rules
fortified by constraints from the classical chiral sum rules. A
careful assessment of the attendant uncertainties is given.
\end{abstract}
\pacs{}

\vspace{1.0in}

\section{Introduction}
In this paper, we present an updated and improved procedure for
obtaining analytical expressions and numerical evaluations of the
matrix elements $\langle (\pi\pi)_{I=2}|{\cal Q}_{7,8}|K^0\rangle$ in
the chiral limit. Recall~\cite{buras1} that the ratio $\epsilon'/\epsilon$ 
can be expressed numerically in terms of operator matrix elements as
evaluated at the scale $\mu = 2~{\rm GeV}$ in the ${\overline {MS}}$-NDR
renormalization scheme~\cite{ciuchini}
\beqa
{\epsilon' \over \epsilon} &=& 20 \times 10^{-4}
\left( { {\cal I}m \lambda_t \over 1.3 \cdot 10^{-3}}\right)
\left[
2.0~{\rm GeV}^{-3} \cdot \langle {\cal Q}_6\rangle^{(0)}_{2~{\rm GeV}}
(1 - \Omega_{\rm IB}) \right. \nonumber \\
& & \left. \phantom{xxxxx} - 0.50~{\rm GeV}^{-3} \cdot 
\langle{\cal Q}_8\rangle^{(2)}_{2~{\rm GeV}} - 0.06 \right]
\label{r1}
\eeqa
where
\beq
\langle {\cal Q}_6\rangle^{(0)} \equiv \langle (\pi\pi)_{I=0} |
{\cal Q}_6 | K^0\rangle \qquad {\rm and} \qquad
\langle {\cal Q}_8 \rangle^{(2)} \equiv
\langle (\pi\pi)_{I=2} | {\cal Q}_8 | K^0\rangle \ \ ,
\label{r2}
\eeq
and all other notation is as in Ref.~\cite{ciuchini}.
In a previous work~\cite{dg}, it was shown that
in the chiral limit the matrix element
$\langle {\cal Q}_8 \rangle_\mu^{(2)}$
(and also $\langle {\cal Q}_7 \rangle_\mu^{(2)}$)
is expressible in terms of certain vacuum matrix
elements\footnote{In this paper, we work with the operator
${\cal Q}_8 \equiv {\bar s}_a\Gamma^\mu_{\rm L}d_b
\left( {\bar u}_b\Gamma_\mu^{\rm R}u_a - {1 \over 2}
{\bar d}_b\Gamma_\mu^{\rm R}d_a - {1 \over 2}
{\bar s}_b \Gamma_\mu^{\rm R}s_a\right)$
to be contrasted with
${\cal Q}_8^{(3/2)} \equiv {\bar s}_a\Gamma^\mu_{\rm L}d_b
\left( {\bar u}_b\Gamma_\mu^{\rm R}u_a -
{\bar d}_b\Gamma_\mu^{\rm R}d_a \right) +
{\bar s}_a \Gamma^\mu_{\rm L}u_b {\bar u}_b\Gamma_\mu^{\rm R}d_a$
as used in Ref.~\cite{dg}. In particular, one has
$\langle (\pi\pi)_{I=2} | {\cal Q}_8^{(3/2)} | K^0\rangle =
2 \langle (\pi\pi)_{I=2} | {\cal Q}_8 | K^0\rangle$. 
Throughout we define $\Gamma^\mu_{{\rm L} \atop {\rm R}} \equiv 
\gamma^\mu (1 \pm \gamma_5)$.}
\beqa
\lim_{p=0}~ \langle (\pi\pi)_{I=2} | {\cal Q}_7
|K^0\rangle_\mu & = & - {2 \over F_\pi^{(0)3}}~\langle {\cal O}_1
\rangle_\mu \ \ , \nonumber \\
\lim_{p=0}~ \langle (\pi\pi)_{I=2} | {\cal Q}_8 |
K^0\rangle_\mu & = & - {2 \over F_\pi^{(0)3}} ~\left[
{1 \over 3} \langle {\cal O}_1 \rangle_\mu +
{1 \over 2} \langle {\cal O}_8 \rangle_\mu \right]
\label{r4}
\eeqa
where $F_\pi^{(0)}$ is the pion decay constant evaluated in the 
chiral limit and the operators ${\cal O}_{1,8}$ are defined as 
\beqa
{\cal O}_1 &\equiv& {\bar q} \gamma_\mu {\tau_3 \over 2} q
~{\bar q} \gamma^\mu {\tau_3 \over 2} q -
{\bar q} \gamma_\mu \gamma_5 {\tau_3 \over 2} q
~{\bar q} \gamma^\mu \gamma_5 {\tau_3 \over 2} q \ \ ,
\nonumber \\
{\cal O}_8 &\equiv& {\bar q} \gamma_\mu \lambda^a
{\tau_3 \over 2} q
~{\bar q} \gamma^\mu \lambda^a {\tau_3 \over 2} q -
{\bar q} \gamma_\mu \gamma_5 \lambda^a {\tau_3 \over 2} q
~{\bar q} \gamma^\mu \gamma_5 \lambda^a {\tau_3 \over 2} q
\ \ .
\label{r6}
\eeqa
In the above, $q = u,d,s$, $\tau_3$ is a Pauli (flavor) matrix,
$\{ \lambda^a \}$ are the Gell~Mann color matrices and the
subscripts on ${\cal O}_1$, ${\cal O}_8$ refer to
the color carried by their currents. Donoghue and Golowich
showed in Ref.~\cite{dg} how to obtain dispersive sum
rules (which we shall refer to as DG1 and DG2 in this paper)
for the vacuum matrix elements of the dimension six operators
${\cal O}_1$ and ${\cal O}_8$.
In a later work~\cite{cdg}, the presence of higher
dimension operators ({\it i.e.} those having dimension $d>6$) was
identified and their impact discussed.

This paper will extend previous work in several significant respects:
\begin{enumerate}

\item We provide a two-loop determination of the dimension-six
contributions to the operator product expansion (OPE) for the isospin
correlator $\Delta \Pi (q^2)$, 
\beqa
& & i \int d^4 x\ e^{i q \cdot x}
\langle 0 |T\left( V^\mu_3  (x) V^\nu_3 (0) -
A^\mu_3 (x) A^\nu_3 (0)\right) | 0 \rangle
\nonumber \\
& & \phantom{xxxxx} = (q^\mu q^\nu - q^2 g^{\mu\nu} )
\Delta  \Pi (q^2) - q^\mu q^\nu \Pi_{A,3}^{(0)}(q^2)
\ \ .
\label{r6.1}
\eeqa 
We give our results in ${\rm {\overline {MS}}}$ renormalization, 
using both NDR and HV schemes for $\gamma_5$, and adopting the same 
evanescent operators scheme as in Refs.~\cite{buras1,buras2,martinelli}.
\item We extend previous work on dispersive sum rules~\cite{dg}
with an updated derivation which takes into account the NLO 
renormalization scheme dependence,  and incorporates 
contributions from so-called higher-dimensional operators~\cite{cdg}.
\item We perform a dispersive evaluation of
$\langle {\cal O}_1 \rangle_\mu$ and $\langle {\cal O}_8 \rangle_\mu$.
Our numerical analysis unifies input from experiment (the existing
data base for the spectral function $\Delta \rho (s)$) with 
rigorous theoretical constraints
embodied by the Weinberg~\cite{sw} and pion mass
difference~\cite{dgmly} chiral sum rules. An important consequence
is the assignment of realistic uncertainties to our results.
\end{enumerate}
\begin{figure}
\vskip .1cm
\hskip 5.0cm
\rotate[l]{
\psfig{figure=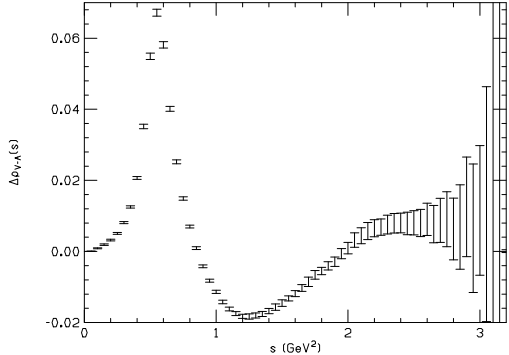,height=2.6in}}
\caption{ALEPH determination of $\Delta \rho (s)$ \hfill
\label{fig:fig0}}
\end{figure}
\section{Theoretical Analysis}
In Ref.~\cite{dg} it was shown how to relate the vacuum matrix elements
of ${\cal O}_1$ and ${\cal O}_8$ to the correlator $\Delta \Pi (q^2)$.
In this section we describe an improved version of the theoretical analysis
of Ref.~\cite{dg}, beginning with some considerations on the central object
of the analysis, the vacuum correlator $\Delta \Pi (q^2)$, defined in
Eq.~(\ref{r6.1}).  Our work exploits  a combination of the two
nonperturbative
representations available for the correlator, the dispersive representation
and the OPE representation at large spacelike momenta.

The dispersive representation reads
\beq
\Delta \Pi (Q^2)
= {1 \over Q^4} \int_0^\infty ds\ {s^2 \over s + Q^2}~
\Delta \rho (s) \ \ ,
\label{r7}
\eeq
where $Q^2 \equiv - q^2$ is the variable for spacelike
momenta and the difference of spectral functions is denoted by
$\Delta \rho (s) \equiv \left[ \rho_{V,3} - \rho_{A,3}\right](s)$.
The ALEPH~\cite{ALEPH} determination of $\Delta 
\rho (s)$ for $4 m_\pi^2 \le s \le m_\tau^2$ 
is displayed in Fig.~\ref{fig:fig0}.

For $Q^2 \gg \Lambda_{\rm QCD}^2$,
$\Delta \Pi (Q^2)$ can be represented via the OPE.
Through ${\cal O} (\alpha_s^2)$ one has
\beq
\Delta \Pi (Q^2) \sim \sum_d \ {1 \over Q^d} \left[
a_d (\mu) + b_d (\mu) \ln{Q^2 \over \mu^2} \right] \qquad
(d = 2,4,6,8,10, \dots) \ \ ,
\label{r11}
\eeq
where $a_d (\mu)$ and $b_d (\mu)$ are combinations of vacuum 
expectation values of local operators of dimension $d$.
We list some properties of the OPE:
\begin{enumerate}
\item For $d \leq 6$, the coefficients $a_d, b_d$ are known 
to $O(\alpha_s^2)$ for $d=2$ and  $O(\alpha_s)$ for $d=4,6$.
\item For $d <6$, $a_d, b_d$ are ${\cal O}(m_q)$ or 
${\cal O}(m^2_q)$ and thus vanish in the chiral limit.
\item $a_6$ and $b_6$ are related to vacuum matrix elements of
the operators ${\cal O}_1$ and ${\cal O}_8$.
\item For $d > 6$, $a_d, b_d$ are partially known (an analysis of $d=8$ can
be
found in Ref.~\cite{bg} for the  vector correlator and Ref.~\cite{ioffe} for
the V-A correlator). In this work their detailed form is not needed.
Hereafter we denote the collective $d>6$ contributions to the OPE as
$\Delta {\overline \Pi} (Q^2)$, that is
\beq
\Delta {\overline \Pi} (Q^2) \sim \sum_{d>6} \ {1 \over Q^d} \left[
a_d (\mu) + b_d (\mu) \ln{Q^2 \over \mu^2} \right] \ \ .
\label{pibar}
\eeq
\end{enumerate}
By virtue of item 3, the $d=6$ OPE coefficients $a_6, b_6$
are of special interest. Here, we consider the 
${\rm {\overline {MS}}}$ renormalization scheme with NDR and HV 
prescriptions for $\gamma_5$ and the evanescent operator basis used 
in Refs.~\cite{buras2,martinelli}. Including ${\cal O} (\alpha_s^2)$ terms 
we find for $N_c = 3$ and $n_f = 3, 4$, 
\beqa
a_6 (\mu) &=& 2 \pi \langle \alpha_s {\cal O}_8 \rangle_\mu +
 A_8 \langle \alpha_s^2 {\cal O}_8 \rangle_\mu +
A_1 \langle \alpha_s^2 {\cal O}_1 \rangle_\mu \ \ ,\nonumber \\
b_6 (\mu) &=& B_8 \langle \alpha_s^2 {\cal O}_8 \rangle_\mu +
B_1 \langle \alpha_s^2 {\cal O}_1 \rangle_\mu \ \ .
\label{r15a}
\eeqa
The coefficients $A_1,A_8$ and $B_1,B_8$ are displayed 
in Table~\ref{tab:scheme}, in terms of their dependence on the 
renormalization scheme and the active number of flavors.
We shall present a derivation of the above results in
Sect.~\ref{OPEmatching}.
%
\begin{table*}[ht]
\caption{Collection of coefficients needed in the calculation at 
various stages.  \label{tab:scheme}}
\vspace{0.4cm}
\begin{center}
\begin{tabular}{c|cc|cc}
& \multicolumn{2}{c}{Three Active Flavors ($n_f = 3$) } &
\multicolumn{2}{c}{Four Active Flavors ($n_f = 4$)} \\ \hline
Scheme & NDR & HV
& NDR & HV \\ \hline
$A_1$ & $2$ & $-10/3$ & $2$ & $-10/3$ \\
$A_8$ & $25/4$ & $21/4$ & $205/36$ & $169/36$ \\ \hline 
$B_1$ & $8/3$ & $8/3$ & $8/3$ & $8/3$ \\
$B_8$ & $-1$ & $-1$ & $-2/3$  & $-2/3$ \\ \hline 
$C_8$ & $-1/6 $ & $11/6$ & $-1/6 $ & $11/6$ \\
$k_s$ & $-1/4 $ & $3/4$ & $-1/4 $ & $3/4$  \\
\end{tabular}
\end{center}
\end{table*}
An additional piece of information which will be needed in our analysis is
the form of $a_6 (\mu)$ at ${\cal O} (\alpha_s)$ in $d= 4 - \epsilon$
dimensions,~\cite{dg}
\beq
a_6 (\mu,\epsilon) = 2 \pi \langle \alpha_s {\cal O}_8 \rangle_\mu \bigg( 1
+ k_s \epsilon \bigg) \ \ .
\label{r15c}
\eeq
The scheme-dependent coefficient $k_s$ can be found in 
Table~\ref{tab:scheme}. 

\subsection{Determination of $a_6 (\mu)$ and $b_6 (\mu)$}
\label{OPEmatching}

In this section we derive the expression for the dimension six terms
in the OPE for $\Delta \Pi (Q^2)$ at NLO in QCD.
We begin by defining an amplitude ${\cal M}$ which describes
the coupling of isospin vector and axialvector currents
via W-boson exchange,
\beq
{\cal M} \equiv {g_2^2 \over 16 F_\pi^2} \int d^4x
\ {\cal D}(x,M_W^2) ~ \langle 0 |T\left( V^\mu_3 (x)
V_{\mu,3} (0) - A^\mu_3 (x) A_{\mu, 3} (0)\right) | 0 \rangle
\ \ .
\label{r8}
\eeq
Our strategy is to perform two different analyses of this
amplitude at NLO in QCD (one of which involves the vacuum correlator
we want to study), and then match the two: consistency then determines
$a_6 (\mu)$ and $b_6 (\mu)$.

Within the first approach, we write~\cite{dg} the amplitude 
${\cal M}$ in the language of an effective theory, involving the
operators ${\cal O}_1$, ${\cal O}_8$ and their
Wilson coefficients $c_1$, $c_8$,
\beq
{\cal M} \simeq {G_F \over 2 \sqrt{2}F_\pi^2}
\left[ c_1 (\mu) \langle {\cal O}_1 \rangle_{\mu}
+ c_8 (\mu) \langle {\cal O}_8 \rangle_{\mu}
\right] \ \ ,
\label{r9}
\eeq
where $\mu$ is the renormalization scale. We define the effective
theory via dimensional regularization within
${\rm {\overline {MS}}}$ renormalization.
The specification of how $\gamma_5$ is treated in $d$-dimensions
(NDR or HV scheme) and of the evanescent operator
basis is needed to uniquely define the effective theory.
We follow here the prescriptions given in Ref.~\cite{buras2}.  The 
Wilson coefficients $c_1(\mu)$ and $c_8(\mu)$ are found by 
performing perturbative matching of the full and effective theories
at scale $\mu = M_{\rm W}$ and then evolving them according
to the renormalization group equations
\beq
\mu {d \over d \mu}~c_k (\mu) = 
\left( \gamma_{\ell k} - 2 \gamma_J  \delta_{\ell k} \right) \, 
c_\ell (\mu)
\ \ .
\label{rg}
\eeq
Here $\{ \gamma_{\ell k} \}$ is the anomalous dimension 
matrix for the operators ${\cal O}_{1,8}$, while $\gamma_J$ 
is the weak current anomalous dimension\footnote{$\gamma_J$ 
is nonzero at two loops in the HV scheme, when subtracting minimally.
In this work we stick to this version of $\overline{\rm MS}-{\rm HV}$ 
renormalization scheme.}. 
For an NLO analysis we need $c_{1,8} (M_W)$ up to non-logarithmic
terms  of order $\alpha_s$, and also the two loop anomalous dimension.
The matching calculation required here has been performed in 
Ref.~\cite{dg} and results in 
\beqa
c_1 (\mu  \simeq M_W) &=&  1 + {\cal O} (\alpha_s^2) \ , 
\ \qquad \ 
c_8 (\mu \simeq M_W) = - \frac{3 \alpha_s (M_W)}{8 \pi} \, \left(
\frac{3}{2} + 2 d_s \right)
\label{wilcoeff}
\eeqa
where $d_s$ is given by
\beq
d_s = \left\{
\begin{array}{cc}
-5/6               & {\rm (NDR)} \nonumber \\
 1/6        & \ \ \ {\rm (HV)} \ \ \nonumber \\
\end{array} \right. \ .
\eeq
The anomalous dimension matrix $\{ \gamma_{\ell k} \}$
is parameterized in terms of the number of active quark flavors
$n_f$ and the number of quark colors $N_c$. At next-to-leading
order (NLO), it has the form
\beq
\gamma_{\rm NLO} = {\alpha_s (\mu) \over 4 \pi} ~\gamma^{(0)} +
\left({\alpha_s (\mu) \over 4 \pi}\right)^2 ~\gamma^{(1)} \ \ .
\label{nlo}
\eeq
The next key observation is that the anomalous dimension matrix
for the operators ${\cal O}_{1,8}$ can be inferred by the
restriction of the full 10$\times$10 matrix of 
Ref.~\cite{buras2,martinelli} to the ${\cal Q}_{7,8}$ subspace.
The two sets of operators  ${\cal O}_{1,8}$
and  ${\cal Q}_{7,8}$ have the same Dirac structure, but differ
in the flavor structure and in the basis for the color structure.
Since only current-current diagrams contribute to the anomalous
dimension matrix of ${\cal O}_{1,8}$, and these contributions are
insensitive to flavor, we only need to handle the
different colour properties. The difference in the color-structure
basis is taken care of by a simple linear transformation.
Denoting by $B^{(0,1)}$ the anomalous dimension matrices of
Refs.~\cite{buras2,martinelli} restricted to  ${\cal Q}_{7,8}$,
those we need for ${\cal O}_{1,8}$ are given by 
\beq
\gamma^{(0,1)} = M \, B^{(0,1)} \, M^{-1} \ ,
\eeq
with
\beq
M = \left( \begin{array}{cc}
  1          &    \ \  0             \nonumber \\
 -2/3        &    \ \ \ \  2          \ \ \nonumber \\
\end{array} \right) \ .
\label{conversion}
\eeq
Using the above ingredients and  the NLO evolution operator
\cite{buras1,martinelli}, we have calculated the Wilson
coefficients in ${\rm {\overline {MS}}}$
renormalization for both NDR and HV schemes, taking
$n_f = 3,4$ and $N_c = 3$. Upon using the expression for the 2-loop
running $\alpha_s$, we have then expanded $c_{1,8}(\mu)$ in powers of
$\alpha_s(\mu)$, finding 
\beqa
c_1 (\mu) &=& 1 + \left( {\alpha_s (\mu) \over \pi}
\right)^2 \left[ {3 A_1 \over 16} \ln {M_W^2 \over \mu^2} +
{3 B_1 \over 32}  \ln^2 {M_W^2 \over \mu^2}
\right] + \dots \ \ ,
\nonumber \\
c_8 (\mu) &=& {\alpha_s (\mu) \over \pi}
\left[ {3 \over 8} \ln {M_W^2 \over \mu^2} -
{3 \over 8} \left( {3 \over 2} + 2 d_s \right)  \right] +
\left( {\alpha_s (\mu) \over \pi}
\right)^2 \left[ {3 A_8 \over 16} \ln {M_W^2 \over \mu^2} +
{3 B_8 \over 32}  \ln^2 {M_W^2 \over \mu^2}
\right] + \dots  ,  \nonumber 
\\
\label{r10}
\eeqa
where $A_1,A_8$ are the scheme-dependent coefficients of 
Table~\ref{tab:scheme}.

The alternate analysis of the amplitude ${\cal M}$ relies on its expression
in terms of the correlator $\Delta \Pi (Q^2)$:
\beq
{\cal M} = {3 G_F M_W^2 \over 32 \sqrt{2}\pi^2
F_\pi^2} \int_{0}^\infty dQ^2 \ {Q^4 \over Q^2 + M_W^2}
\Delta \Pi (Q^2) \ \ .
\label{r10.5}
\eeq
We partition the amplitude ${\cal M}$ as
\beq
{\cal M} \ = \ {\cal M}_< (\mu) \ + \ {\cal M}_> (\mu)
\label{r12a}
\eeq
where the  component ${\cal M}_< (\mu)$ arises from contributions with
$Q<\mu$,
\beq
{\cal M}_< (\mu) = {3 G_F \over 32 \sqrt{2}\pi^2
F_\pi^2} \int_0^{\mu^2} dQ^2 \ Q^4 ~
\Delta \Pi (Q^2) + {\cal O}(\mu^2 / M_W^2)
\label{r13}
\eeq
and the component ${\cal M}_> (\mu)$ contains the
 contributions with $Q>\mu$,
\beq
{\cal M}_>(\mu) = {3 G_F M_W^2 \over 32 \sqrt{2}\pi^2
F_\pi^2} \int_{\mu^2}^\infty dQ^2 \ {Q^4 \over Q^2 + M_W^2}
\Delta \Pi (Q^2) \ \ .
\label{r14}
\eeq
Employing the OPE for $Q^6 \Delta \Pi (Q^2)$ in
${\cal M}_>(\mu)$ and performing the $Q^2$ integration,
one can compare this representation
with Eqs.~(\ref{r9})-(\ref{r10}).
Requiring the consistency of the two approaches yields
the expressions of Eq.~(\ref{r15a}) for $a_6 (\mu)$ and
$b_6 (\mu)$, as well as the sum rule DG1 
({\it cf.} Eq.~(\ref{r16}) below), 
to be derived in a different way in the next section. 
\subsection{The Two DG Sum Rules}
In this section we present a short derivation of the two DG
sum rules \cite{dg}, including effects which were neglected in the
original derivation.  Let us start with the first sum rule. 
Consider the V-A correlator in coordinate space for 
$x \rightarrow 0$, regularized dimensionally 
\beqa
\langle {\cal O}_1 \rangle_{\mu} &\equiv&
\langle 0 |T\left( V^\mu_3  (0) V_{\mu,3}  (0) -
A^\mu_3 (0) A_{\mu, 3} (0)\right) | 0 \rangle_\mu
\nonumber \\
&=& { (d - 1) \mu^{4 -d}
\over (4 \pi)^{d/2} \Gamma(d/2)} \int_0^\infty dQ^2
\ ~Q^d  \Delta \Pi (Q^2) \ \ .
\label{dgsr1}
\eeqa
Let us now break up the above integral at a scale $\mu^2$, within the
region of applicability of perturbative QCD. The integration
over $Q^2 < \mu^2$ is UV finite and can be performed in $d=4$. The
integration
for $Q^2 > \mu^2$ generates UV singularities. Using the OPE
representation for $\Delta \Pi (Q^2)$ in this region, one sees 
that the UV divergence is
related to the $Q^{-6}$ term in the expansion, while $\Delta {\overline \Pi}
(Q^2)$ leads to an UV finite term. Employing Eq.~(\ref{r15c}) for
$a_6 (\mu,\epsilon)$, performing the $Q^2$ integration and subtracting the
$1/\epsilon$ pole according to the $\overline{\rm MS}$ scheme, we
obtain 
\beq
\langle {\cal O}_1 \rangle_\mu -
{3 C_8 \over 8 \pi}
\langle \alpha_s {\cal O}_8 \rangle_\mu
= \bar{I}_1 (\mu) 
\qquad {\rm (DG1)}  
\label{r16}
\eeq
where the scheme-dependent coefficient $C_8$ can be found in 
Table~\ref{tab:scheme},\footnote{Although we have employed 
three distinct quantities $k_s, d_s, C_8$ to indicate how 
scheme-dependence enters in different parts of the analysis, 
they are in fact related by $C_8 = 1/3 + 2 k_s = 3/2 + 2 d_s$.}  
and $\bar{I}_1 (\mu)$ is defined in terms of the 
dispersive integrals  $I_1 (\mu)$, $H_1 (\mu)$ as follows
\beq
\bar{I}_1 (\mu) = 
{3 \over (4 \pi)^2} \left[ I_1 (\mu) + H_1 (\mu) \right] \ \ , 
\label{i1def}
\eeq
where 
\beqa
I_1 (\mu) & \equiv &  \int_0^\infty ds\ s^2 \ln
\left({s + \mu^2 \over s} \right) ~\Delta\rho(s)
\equiv \int_0^{\mu^2} dQ^2 \ Q^4 \Delta\Pi(Q^2) \ \ , 
\label{vcin1} \\
H_1 (\mu) & \equiv &
\int_{\mu^2}^\infty dQ^2 \ Q^4~\Delta{\overline \Pi} (Q^2)
\ \ . \label{r17}  
\eeqa

Evaluating the OPE of Eq.~(\ref{r11}) at the
point $Q = \mu$ yields the second DG sum rule,~\cite{dg}
\beq
2 \pi \langle \alpha_s {\cal O}_8
\rangle_\mu + A_1 \langle \alpha_s^2
{\cal O}_1 \rangle_\mu +  A_8 \langle \alpha_s^2
{\cal O}_8 \rangle_\mu 
= 2 \pi \alpha_s (\mu)  \bar{I}_8 (\mu) 
\qquad ({\rm DG2}) 
\label{va38}
\eeq
where we have made use of Eq.~(\ref{r15a}) and define
$\bar{I}_8 (\mu)$ as
\beq
\bar{I}_8 (\mu) = \frac{1}{2 \pi \alpha_s (\mu)} 
\bigg[ I_8 (\mu) - H_8 (\mu) \bigg] \ \ , 
\label{i8def}
\eeq
where 
\beqa
I_8 (\mu) & \equiv &  \int_0^\infty ds\ s^2 {\mu^2 \over s + \mu^2}
~\Delta\rho (s) = \mu^6 \Delta\Pi (\mu)  \ \ , \label{vcin2} \\
H_8 (\mu) & \equiv & \mu^6 \Delta {\overline \Pi} (\mu) \ \ . 
\label{r19} 
\eeqa
In Ref.~\cite{dg} the terms $H_{1,8} (\mu)$, subleading at high $\mu$,
were neglected. They encode the effect of higher dimensional 
operators and lead to potentially large effects \cite{cdg,knecht}.

We can summarize the work thus far via the linear relations
\beq
\left(
\begin{array}{cc}
1 \ \  & - 3 C_8 \alpha_s (\mu) / 8 \pi \\
\  \ \ & \ \  \\
A_1 \alpha_s (\mu) / 2 \pi \ \ & 
\left( 1 +  A_8 \alpha_s (\mu) /2\pi  \right) 
\end{array}
\right) \ \left(
\begin{array}{c}
\langle {\cal O}_1 \rangle_\mu \\
\ \ \ \\
\langle {\cal O}_8 \rangle_\mu 
\end{array} 
\right) = 
\left(
\begin{array}{c}
\bar{I}_1 (\mu)  \\
\ \ \ \\
\bar{I}_8 (\mu) 
\end{array} 
\right) \ , 
\label{matrix}
\eeq
which allow us to keep track of the scheme dependence of the 
matrix elements at NLO. 
As a check on our calculation of the coefficients $A_{1,8}, C_8$, 
we can derive the relation between the HV and NDR matrix elements 
of ${\cal Q}_{7,8}$. To do so, we need to solve Eq.~(\ref{matrix}) 
for $\langle {\cal O}_{1,8} \rangle_{\mu}$ in both schemes and then 
convert to ${\cal Q}_{7,8}$ using the matrix $M$ of 
Eq.~(\ref{conversion}). 
We find
\beq
\left(
\begin{array}{c}
\langle {\cal Q}_7 \rangle_\mu \\
\ \ \ \\
\langle {\cal Q}_8 \rangle_\mu 
\end{array} 
\right)^{HV} =
\left(
\begin{array}{c}
\langle {\cal Q}_7 \rangle_\mu \\
\ \ \ \\
\langle {\cal Q}_8 \rangle_\mu 
\end{array} 
\right)^{NDR}
\ + \ \frac{\alpha_s (\mu) }{\pi} 
\left(
\begin{array}{cc}
-1/2 \ \  &   3/2 \\
\  \ \ & \ \  \\
1 & 1 
\end{array}
\right) \ 
\left(
\begin{array}{c}
\langle {\cal Q}_7 \rangle_\mu \\
\ \ \ \\
\langle {\cal Q}_8 \rangle_\mu 
\end{array} 
\right)^{NDR} \ , 
\label{matrix1}
\eeq 
in agreement with the result given in Refs.~\cite{buras2,buras3}.

\subsection{Additional Comments} 
Let us take note of the differences of the preceding analysis from 
that in the previous
work by two of the present authors~\cite{dg}. In the earlier work, the
information about perturbative corrections in the NDR scheme was taken
from the paper of Lanin {\it et al.}~\cite{Lanin} in the literature 
on QCD sum rules. This is now seen to differ in the renormalization 
conventions and
use of evanescent operators from the choices made in the two loop
analysis of the weak Hamiltonian. In the present work we have adopted
the conventions of the weak interaction studies. One fortunate benefit
from this change is that the results are less sensitive to higher order
perturbative corrections than in the previous analysis. In the earlier
work we also did not present all results for the HV scheme, whereas 
this is accomplished in the present study. Finally, while the previous work
led directly to the uncovering of the effects of higher dimension
operators, these effects were not explicitly incorporated into the
formulas or numerical work of that paper. In the present paper we have
explicitly included these effects.

\begin{figure}
\vskip .1cm
\hskip 2.5cm
\psfig{figure=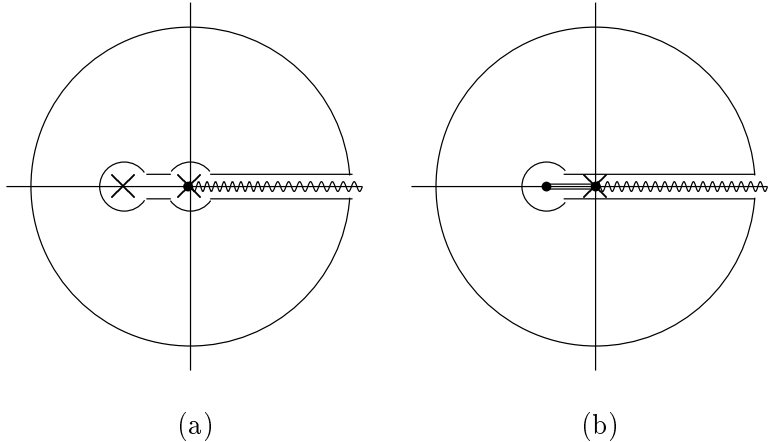,height=2.3in}
\caption{Integration contours for (a) $I_8$, (b) $I_1$ \hfill
\label{fig:fig1}}
\end{figure}

The integrals $I_8 (\mu)$ and $I_1(\mu)$ will be 
central to our analysis. They are related by
\beq
I_8 (\mu) = \mu^2 {d \over d \mu^2} I_1 (\mu) \ \ ,
\label{relate}
\eeq
and with the aid of the Weinberg sum rules~\cite{sw}, it is possible
to express $I_8 (\mu)$ in several equivalent forms, {\it e.g.}
\beq
I_8 (\mu) = - \mu^4 \int_0^\infty ds\ {s \over s + \mu^2}
~\Delta\rho (s) \ \ .
\label{equiv}
\eeq
Moreover, there is an alternative pathway to 
Eqs.~(\ref{vcin1}),(\ref{vcin2}).  One simply evaluates 
two contour integrals in the complex $s \equiv q^2$ plane, 
\beq
\oint_{\Gamma_{k}} \ ds \ K_k (s,\mu ) ~\Delta \Pi (s) = 0 
\qquad (k = 1,8) \ \ , 
\label{contour}
\eeq
where the weights $K_k (s, \mu)$ are 
\beq
K_1 (s,\mu) \equiv s^2 \ln{s + \mu^2 \over s} \ , 
\qquad 
K_8 (s,\mu) \equiv {\mu^2 s^2 \over s + \mu^2} \ \ .
\label{kernel}
\eeq
The contours $\Gamma_{8}$, $\Gamma_{1}$ are shown 
in Figs.~\ref{fig:fig1}(a),(b), with the understanding 
that the radius of each outer circle is to be taken to infinity.  
The unitarity cut in Figs.~\ref{fig:fig1}(a),(b) is denoted by 
a wiggly line along the positive real axis.  The pole in 
$K_8 (s,\mu)$ at $s = - \mu^2$ is denoted in Fig.~\ref{fig:fig1}(a) 
by the cross on the negative $s$ axis. The cut 
in $K_1 (s, \mu)$ occurring for $-\mu^2 \le s \le 0$ is 
denoted by the bold straight line.

Obtaining numerical values (along with meaningful
error estimates) for $I_8 (\mu)$ and $I_1(\mu)$ turns out
to be a highly nontrivial task. It is this problem that
we turn to in the next section.

\section{Numerical Analysis}
The goal of this section is to arrive at numerical estimates for the
input vector on the right hand side of Eq.~(\ref{matrix}).
We shall accomplish this by means of a constrained evaluation of
the dispersive integrals $I_{1,8} (\mu)$ using an
approach which we call the {\it residual weight approximation}.
This determination can be applied over a range of possible
scales, {\it e.g.} $2 \le \mu {\rm (GeV)} \le 4$. In fact,
we choose a scale $\mu = 4$~GeV sufficiently large for
higher-dimension effects to be negligible and then employ a
two-loop renormalization group analysis to evolve our result
down to scale $\mu = 2$~GeV.  The decision to set $H_{1,8} (4) = 0$ 
is supported by a preliminary study using Finite Energy Sum 
Rules (FESR)~\cite{budapest}.  Work on this topic continues and 
will be reported on in a separate publication~\cite{cdgm6}.
\subsection{Residual Weight Approximation}
Direct evaluation of $I_1 (\mu)$ and $I_8 (\mu)$ via
Eqs.~(\ref{vcin1}) and (\ref{vcin2}) is impossible since 
data for the spectral function $\Delta \rho (s)$ exists
only in the interval $4 m_\pi^2 \le s \le m_\tau^2$.
Even direct integration of the contribution from the
region $s<m_\tau^2$ is potentially problematic since
strong cancellations significantly enhance the impact of
experimental errors. This is compounded by the
fact that the weights $K_{1,8}(s,\mu)$ are strongly increasing
with $s$ and hence weight the high-$s$ region, where data errors are
large, more strongly than the low-$s$, low error region.
As an example, defining $I_{1,8}(s_0,\mu)$
to be versions of Eqs.~(\ref{vcin1}) and (\ref{vcin2}) containing
spectral contributions only up to $s=s_0$, the errors on
$I_{1,8}(s_0,\mu)$, determined using the ALEPH
covariance matrix, already exceed $50\%$ of our final central values
for $s_0\sim 2.1\ {\rm GeV}^2$. The possibility of obtaining
reliable determinations of $I_{1,8}(\mu)$ thus depends
critically on our ability to impose additional constraints
on these integrals.


Although $\Delta\rho (s)$ is not known above $s=m_\tau^2$, and is
determined with insufficient precision for our purposes between
$2\ {\rm GeV}^2$ and $m_\tau^2$, indirect information on its
behavior in these regions, in the chiral limit, is provided by
the two Weinberg sum rules~\cite{sw} 
\beq
W1 \equiv \int_0^\infty ds\
\Delta\rho(s) = F_\pi^{(0)2} \ ,
\qquad
W2 \equiv \int_0^\infty ds\ s ~\Delta\rho(s) = 0 \ \ ,
\label{ch1}
\eeq
where $F_\pi^{(0)}$ is the pion decay constant
in the chiral limit, and the sum rule for the EM pion mass
splitting~\cite{dgmly}
\beq
W3 \equiv \int_0^\infty ds\ s \ln {s \over \Lambda^2}
~\Delta\rho(s) = - {F_\pi^{(0)2}} {4 \pi \over 3 \alpha}
\Delta m^{(0)2}_\pi \ \ ,
\label{ch2}
\eeq
where $\Delta m^{(0)2}_\pi$ is the pion squared-mass splitting 
in the chiral limit,
$\Delta m^{(0)2}_\pi\simeq m_{\pi^\pm}^2 - m_{\pi^0}^2$,
and we will take $\Lambda =1\ {\rm GeV}$, for definiteness, in
what follows (the LHS of Eq.~(\ref{ch2}) is independent of $\Lambda$).
An additional constraint is provided by the asymptotic OPE
form for $\Delta\rho (s)$,
\beq
\Delta \rho (s) \sim {1 \over s^3}
\left[ B_1  \langle \alpha_s^2
{\cal O}_1 \rangle_\mu
+ B_8 \langle \alpha_s^2
{\cal O}_8 \rangle_\mu \right] + \dots \ \ ,
\label{va8}
\eeq
valid for sufficiently large $s$, say $s>s_A$.

Because of the asymptotic constraint of Eq.~(\ref{va8}),
uncertainties in the evaluation of any spectral integral
are dominated by contributions from
the region $s\sim 2-2.5\ {\rm GeV}\rightarrow s_A$,
for which spectral data is absent, or has large
errors. Letting $K(s,\mu)$ stand for
either $K_1(s,\mu)$ or $K_8(s,\mu)$,
an obvious way to take advantage of the constraints
provided by the chiral sum rules is to write $K(s,\mu)$ in the form
\beq
K (s,\mu) = C (s,\mu) + \Delta K (s,\mu) \ \ ,
\label{ch3}
\eeq
where $C(s,\mu)$ is an arbitrary linear combination of the weights
occurring in the chiral sum rules W1, W2 and W3,
\begin{equation}
C (s,\mu ) \equiv x + y s +
z s \ln \left({s \over \Lambda^2}\right) \ \ ,
\label{ch4}
\eeq
and the `residual weight', $\Delta K(s,\mu)$, is defined trivially by
\beq
\Delta K (s,\mu) \equiv K (s,\mu) - C (s,\mu) \ \ .
\label{ch3z}
\eeq
The residual weight representation for $K(s,\mu)$ generates
analogous representations for the integrals $I \equiv I_{1,8}$,
\beq
I (\mu) = I_{\rm chiral} (\mu) + \Delta I (\mu) \ \ ,
\label{ch4a}
\eeq
where
\beq
I_{\rm chiral}(\mu) \equiv \int_0^\infty\, ds\, C(s,\mu)\Delta\rho (s)
= F_\pi^{(0)2} \left[ x - z \left({4 \pi \over 3 \alpha}\right)
\Delta m^{(0)2}_\pi \right] \ \ ,
\label{ch5}
\eeq
and
\begin{equation}
\Delta I(\mu)\equiv \int_{0}^{\infty} \, ds\, 
\Delta K(s,\mu)\Delta\rho (s)\ \ .
\end{equation}
Errors on the determination of $I_{1,8}(\mu)$ associated
with uncertainties in our knowledge of $\Delta\rho (s)$
can then be reduced by adjusting the free parameters $x,y,z$
so as to make $\Delta K(s,\mu)$ small in the region between
$\sim 2.5\ {\rm GeV}^2$ and $s_A$. We will refer to this
region as the ``matching region'' in what follows.

In order to choose $x,y,z$ in such a way as to minimize
the total errors on $I_{1,8}$, it is necessary to take into
account the uncertainties in our knowledge of the chiral limit
values, $F_\pi^{(0)}$ and $\Delta {m_\pi^{(0)2}}$, which
enter the W1 and W3 chiral sum rules.
Taking as input the values~\cite{abt}
\beq
F_\pi^{(0)} = (0.0871 \pm 0.0026)~{\rm GeV} \qquad {\rm and} \qquad
\Delta m^{(0)2}_\pi = (0.001174 \pm 0.000055)~{\rm GeV}^2 \ \ ,
\label{ch51}
\eeq
the error, $E_{\rm chiral}$, on $I_{\rm chiral}(\mu)$ becomes
\beq
E_{\rm chiral} =
\pm 0.000453\ {\rm GeV}^2~| x - 0.674\ {\rm GeV}^2~z|
\pm 0.000240\ {\rm GeV}^4~z \ \ .
\label{ch12}
\eeq
Based on the NNLO chiral expansion for $F_\pi$ given in 
Ref.~\cite{abt}, we believe that the difference between the physical 
and NNLO chiral values represents an extremely conservative estimate 
of the uncertainty on $F_\pi^{(0)}$.  We have therefore taken half this 
value ($\pm 0.0026$) as the error cited above in an attempt to employ a 
`one-sigma' theoretical error akin to that used in assigning 
experimental error. 

The corresponding $x,y,z$-dependent error on
$\Delta I(\mu)$ is obtained as follows.
We first partition the range of integration into three intervals,
\beqa
\Delta I &=& [\Delta I]_{\rm data} + [\Delta I]_{\rm int} +
[\Delta I]_{\rm asy} \ \ ,
\nonumber \\
&=& \left[ \int_0^{m_\tau^2} + \int_{m_\tau^2}^{s_A} +
\int_{s_A}^\infty
\right] \ ds \ \Delta K (s,\mu) ~\Delta\rho(s) \ \ ,
\label{ch6a}
\eeqa
where, as above, $s_A$ represents the point beyond which one can
employ the OPE representation of $\Delta\rho (s)$.
Our results are insensitive to the actual choice of $s_A$ but
for definiteness we work with $s_A = 5~{\rm GeV}^2$.
The three regions (`data', `intermediate' and `asymptotic')
are identified by the different ways in which the spectral
function $\Delta\rho (s)$ is treated.
For $[\Delta I]_{\rm data}$, the compilation
of $\Delta\rho(s)$ provided by ALEPH is used. In practice,
we sum over those experimental bins covering the range
$s \leq 3.15~{\rm GeV}^2$. For $[\Delta I]_{\rm int}$, we assume
$|\Delta\rho(s)| < \Delta \rho_{\rm max} = 0.005$ throughout the 
interval.  Since the three known peaks in $\Delta\rho(s)$ decrease 
in magnitude with increasing $s$ (as expected given the $1/s^3$ 
asymptotic fall-off) and since $\Delta\rho(s) \simeq 0.005$ for 
$s \simeq m_\tau^2$, this seems to us a reasonable bound.  
The treatment of $[\Delta I]_{\rm int}$ is discussed
in more detail below. 
For $[\Delta I]_{\rm asy}$, we take the form given in
Eq.~(\ref{va8}) but with a numerical value assumed for the
numerator, {\it viz} $\Delta\rho(s) \sim 7.3 \cdot 
10^{-5}~{\rm GeV}^6/s^3$,
which is compatible with our final results. The error on
$\Delta I(\mu)$ is obtained by adding the errors
associated with each of these three contributions in quadrature.
The determination of the individual error contributions is
discussed below.

The source of the uncertainty on $[\Delta I]_{\rm data}$
is $\Delta\rho(s)$, the most important component being that
due to the experimental errors determined
by ALEPH as part of their extraction of $\Delta\rho (s)$
from hadronic $\tau$ decay data. For each weighted spectral
integral, this error is computed using the ALEPH
covariance matrix. A much less significant uncertainty arises from
the fact that our theoretical analysis corresponds to the chiral world,
whereas our data sample is taken in the physical world. It is argued in
Ref.~\cite{dg0} that such an effect is anticipated to be small,
${\cal O}(m_\pi^2 /m_\rho^2)$.  This expectation is borne out in 
the analysis of Ref.~\cite{moussallam}.

For the intermediate contribution $[\Delta I]_{\rm int}$,
we adopt the strategy of assigning the value
\beq
[\Delta I]_{\rm int} = 0 \pm E_{\rm int} \ \ ,
\label{ch13}
\eeq
where $E_{\rm int}$ is determined by using the Cauchy-Schwarz 
inequality,
\beq
|\Delta I_{\rm int}| < E_{\rm int} \equiv 
|\Delta\rho_{\rm max}| \sqrt{{s_A} - m_\tau^2}
\cdot \sqrt{ \int_{m_\tau^2}^{s_A} ds\, \left[\Delta K(s,\mu)
\right]^2} \ \ .
\label{ch14}
\eeq
We recall that the quite reasonable bound
$|\Delta\rho_{\rm max} | =  0.005$ is employed throughout
the intermediate region.

For the asymptotic contribution $[\Delta I]_{\rm asy}$,
we assume a $100\%$ uncertainty.
Although this assignment of error is arbitrary,
the asymptotic contribution turns out to be so
tiny that the $100\%$ error does not affect
the overall error-minimization in any way.

At this point, we have a residual weight representation
for the integral $I$ ($I_1$ or $I_8$) parameterized in terms
of three constants $x,y,z$. Since $E_{\rm chiral}$
is independent of $y$, we may reduce $E_{\rm int}$
without adversely affecting $E_{\rm chiral}$,
for any $x,z$, by choosing $y$ so as to appropriately minimize
$\vert \Delta K(s,\mu)\vert$ over the matching region.
The lower edge of this region is chosen to be
$s_0=2.5\ {\rm GeV}^2$, rather than $m_\tau^2$, in order that
contributions from that portion of the spectrum
where experimental errors are large will be suppressed
once one has performed this minimization.
We thus fix $y=y(x,z)$ by minimizing
\begin{equation}
\int_{s_0}^{s_A} ds \ \left[ K(s,\mu) - x - y s -
z s \ln \left({s \over \Lambda^2}\right) \right]^2 \ \ .
\label{ch14a}
\eeq
Like $E_{\rm chiral}$, the errors on the three contributions to
$\Delta I(\mu)$ now depend only on $x$ and $z$, so we may
combine all errors in quadrature, and minimize the total
error with respect to $x,z$.

The results of the optimized versions of the $I_{1,8}(\mu)$
residual weight analyses are
\beq
\begin{array}{c|c|c}

\mu~{\rm GeV} & I_1(\mu)~({\rm GeV}^6) & I_8(\mu)~({\rm GeV}^6) \\
\hline
2 & - (39.7 \pm 3.1) \cdot 10^{-4} & - (26.2 \pm 3.0) \cdot 10^{-4} \\
3 & - (63.1 \pm 5.9) \cdot 10^{-4} & - (31.4 \pm 5.9) \cdot 10^{-4} \\
4 & - (82.1 \pm 9.3) \cdot 10^{-4} & - (34.4 \pm 9.4)  \cdot 10^{-4}
\end{array}
\label{numvalues}
\eeq
The uncertainty in each case is acceptable.

Let us study the various contributions to the $I_{1,8}$ integrals 
at some scale, say, $\mu = 4$~GeV.  For each integral $I$, there will be 
four contributing sources:  $I_{\rm chiral}$, $\Delta I_{\rm data}$,   
$\Delta I_{\rm interm}$ and $\Delta I_{\rm asymp}$.  Each source 
will have an associated uncertainty, except for the chiral 
contribution where we list {\it two} errors, the 
first from $F_\pi^{(0)}$ and the second from 
$\Delta m^{(0)2}_\pi$.  The numerics are as follows:
\beq
\begin{array}{c|c|c}

{\rm Source} & I_1(4)~({\rm GeV}^6) & I_8(4)~({\rm GeV}^6) \\
\hline
{\rm Chiral} & - (23.9 \pm 1.4 \pm 9.1) \cdot 10^{-4} & (24.3
 \pm 1.5 \pm 7.5) \cdot 10^{-4} \\
{\rm Data} & - (58.1 \pm 0.9) \cdot 10^{-4} & - (58.8 \pm 3.2) \cdot 10^{-4} \\
{\rm Intermed.} & (0 \pm 0.7) \cdot 10^{-4} & (0 \pm 4.6) \cdot 10^{-4} \\
{\rm Asympt.} & - (0.1 \pm 0.1) \cdot 10^{-4} & (0.1 \pm 0.1) \cdot 10^{-4} \\
\end{array}
\label{numvalues2}
\eeq
The largest source of uncertainty in our results turns out to originate 
with the $4.6\%$ error in Eq.~(\ref{ch51}) for the pion squared-mass 
splitting (equivalent to $\Delta m_\pi^{\rm (QCD)} = 0.32 +/- 
0.20$~MeV).~\cite{abt}  Our knowledge of this $(m_u-m_d)^2$ effect 
arises mostly from studying the kaon mass difference. 
There, the effect of electromagnetism 
(which shows significant SU(3) breaking) 
is subtracted and the remaining kaon mass difference fit 
by $m_d - m_u$ in lowest order chiral perturbation theory. 
However, the next order chiral
corrections could easily be sizeable at the kaon mass scale.  
and one must also cope with 
$\pi^0$-($\eta, \eta'$) mixing.  In light of these effects, 
we feel the current $4.6\%$ error cited in Ref.~\cite{abt} is realistic.  

Finally, we point out that a possible modification of the 
procedure described 
above would be to include the $F_\pi^{(0)3}$ factors of 
Eq.~(\ref{r4}) {\it ab initio} in our error minimization 
analysis.  In other words, we would be determining integrals 
$J_{1,8} \equiv I_{1,8}/F_\pi^{(0)3}$ rather than just 
$I_{1,8}$.  Such a program produces results completely 
compatible with those of Eq.~(\ref{numvalues}) and with no 
significant reduction in the size of the errors.

\subsection{Renormalization Group Evolution and Results}

In this section we outline the procedure for arriving at the
physical matrix elements $\langle (\pi\pi)_{I=2} | {\cal Q}_{7,8}
|K^0\rangle$ in the chiral limit.  

Using the values for $I_{1,8} (\mu = 4~{\rm GeV})$ obtained in the previous
section and setting $H_{1,8} (\mu = 4~{\rm GeV}) = 0$, we solve
Eq.~(\ref{matrix}) for $\langle {\cal O}_{1,8} \rangle_{\mu=4~{\rm GeV}}$.
Since we work at the scale $\mu = 4$ GeV we use the values of our
coefficients corresponding to four active flavors ($n_f =
4$). Moreover we use $\Lambda_{QCD}^{n_f = 4} = (340 \pm 50)$ MeV.
We subsequently apply the NLO renormalization group evolution, 
to obtain $\langle {\cal O}_{1,8} \rangle_{\mu=2~{\rm GeV}}$. Denoting by 
$U(\mu_L, \mu_H)$ the operator governing the NLO evolution of the 
Wilson coefficients $(c_{1} (\mu),c_{8} (\mu)) =  \vec{c}^T (\mu)$ 
between the scales $\mu_L$ and $\mu_H$, 
\beq
\vec{c} (\mu_L) = U(\mu_L, \mu_H) \ \vec{c} (\mu_H) \ , 
\eeq
the local operators $({\cal O}_{1} (\mu),{\cal O}_{8} (\mu)) =
\vec{{\cal O}}^T (\mu)$ evolve according to:
\beq
\langle \vec{{\cal O}} \rangle (\mu_L) = \bigg[ U^{T} (\mu_L,\mu_H) 
\bigg]^{-1} \ \langle \vec{{\cal O}} \rangle (\mu_H) \ . 
\eeq
The results for $\langle {\cal O}_{1,8} \rangle_\mu$ 
and $\langle (\pi\pi)_{I=2} | {\cal Q}_{7,8} |K^0\rangle_\mu$ 
are reported in Table~\ref{tab:Ois} for the scales 
$\mu = 2,4$~GeV.  
The vacuum matrix elements $\langle {\cal O}_{1,8} \rangle_\mu$ 
are converted to the $K \rightarrow \pi \pi$ 
matrix elements through Eqs.~(\ref{r4}).  We have added in quadrature 
the uncertainties from the vacuum matrix elements and from the 
$1/F_\pi^{(0)3}$ factor.  

\begin{table*}[t]
\caption{Numerical values at $\mu = 2$~GeV and $\mu = 4$~GeV.\label{tab:Ois}}
\vspace{0.4cm}
\begin{center}
\begin{tabular}{c|cc|cc}
& \multicolumn{2}{c}{$\langle {\cal O}_{1} \rangle_\mu ({\rm GeV}^6)$} & 
  \multicolumn{2}{c}{$\langle {\cal O}_{8} \rangle_\mu  ({\rm GeV}^6)$} 
\\ \hline 
$\mu$~(GeV) & NDR & HV
& NDR & HV \\ \hline
$2$ & $- (0.53 \pm 0.34) \cdot 10^{-4}$ & 
$- (1.64 \pm 0.18 ) \cdot 10^{-4}$ 
& $- (14.4 \pm 4.3) \cdot 10^{-4}$ &
$- (15.2 \pm 4.4) \cdot 10^{-4}$ \\
$4$ & $- (1.47 \pm 0.18) \cdot 10^{-4}$ & 
$- (2.59 \pm 0.33) \cdot 10^{-4}$ &
$- (19.3 \pm 5.4) \cdot 10^{-4}$ &
$- (20.3 \pm 5.6) \cdot 10^{-4}$ 
\\ \hline \hline
& \multicolumn{2}{c}{$\langle (\pi\pi)_{I=2} | {\cal Q}_7 |
K^0\rangle_\mu~({\rm GeV}^3)$} &
\multicolumn{2}{c}{$\langle (\pi\pi)_{I=2} | {\cal Q}_8 |
K^0\rangle_\mu~({\rm GeV}^3)$} \\ \hline 
$\mu$~(GeV) & NDR & HV
& NDR & HV \\ \hline
$2$ & $0.16 \pm 0.10$ & $0.49 \pm 0.07$
& $2.22 \pm 0.67$ & $2.46 \pm 0.70$ \\
$4$ & $0.44 \pm 0.07$ & $0.78 \pm 0.12$
& $3.06 \pm 0.87$ & $3.32 \pm 0.90$ \\ 
\end{tabular}
\end{center}
\end{table*}

\section{Final Comments}
Our goals in this paper have been to exactly match the dispersive
description of the weak amplitudes $\langle (\pi\pi)_{I=2}|{\cal
Q}_{7,8}|K^0\rangle$ in the chiral limit to the specific conventions 
of the existing two-loop calculations of the Wilson 
coefficients which appear in the effective weak hamiltonian, and to
provide a numerical evaluation using as input only experimental 
data and the rigorous theoretical constraints of QCD.

In order to accomplish the matching, we adopted the operator basis
and renormalization conventions of Ref.~\cite{buras2}. On one hand we
analyze the OPE by matching the hamiltonian at the scale $M_{\rm W}$ 
and evolving the operators to a lower scale $\mu$. By simultaneously
considering the dispersive representation of the matrix element, we use
the OPE to identify the ${\overline {\rm MS}}$ operators at the 
scale $\mu$ in both the NDR and HV schemes. Our results also 
identify the role of higher dimensional operators, which enter 
the dimensionally regularized matrix elements.

In the numerical analysis we have developed the `residual weight' 
method, which provides an evaluation requiring as input only 
data and the chiral-limit values $F_\pi^{(0)}$ and 
$\Delta m_\pi^{(0)}$. This method expresses the dispersive integrals 
$I_1 (\mu)$ and $I_8 (\mu)$ in terms of known quantities (the other 
known dispersive sum rules) plus a residual integral. For the 
residual integral, we make use of the ALEPH data and employ 
only a conservative bound on the spectral function in the region 
where $\tau$ decay data is not available.  We have
chosen to evaluate $I_1 (\mu)$ and $I_8 (\mu)$ 
at large values of $\mu$ where the effects of higher
dimension operators are negligible.  We then use the renormalization
group to obtain the values of the matrix elements at other scales

Our results lead to the matrix elements given in Table~II and Table~III.
Because we only use real experimental input, in combination with 
very conservative assumptions about the uncertainties on 
$F_\pi^{(0)}$ and $\Delta m_\pi^{(0)}$, the final errors we 
quote are very conservative ones.  
These are the maximal error bars of the dispersive evaluation
--- our method has been designed to give as pure and conservative an
evaluation from data as possible, and other methods may reduce the error
bars. For example, Eq.~(\ref{numvalues}) shows that the uncertainty 
on the residual
weight analysis decreases dramatically at lower values of $\mu$. To
directly work at $\mu = 2$~GeV one needs to evaluate, and correct for,
the effects of higher dimension operators. We have been studying this
problem through the use of FESR and preliminary
indications yield results consistent with the present evaluation but
with reduced error bars~\cite{budapest}. However that analysis requires
the development of further techniques and consistency checks, and we
will present it in detail elsewhere~\cite{cdgm6}. 
The present method stands on its
own as being especially simple and model independent.

In Table~\ref{tab:compare}, we compare our results at the scale 
$\mu = 2$~GeV to other determinations: 
the lattice Monte Carlo simulation~\cite{donini}, 
the large $N_c$ approach~\cite{knecht}, 
the so-called X-boson approach~\cite{lund} 
and the vacuum saturation approximation (VSA). 
We make the following remarks regarding this comparison.

The Monte Carlo simulation of Ref.~\cite{donini} is displayed 
in the `Lattice' row of Table~\ref{tab:compare}.  Note that 
their HV results have been converted to the HV scheme used 
in this work. As regards comparison between the lattice results 
and ours, we should point out that the two approaches correspond to 
somewhat different limits. The lattice evaluation 
extrapolates the results to the physical values of the quark masses,
while our dispersive analysis applies to the amplitude in the chiral
limit. Even so, the extrapolation between the chiral limit and the
physical masses has been addressed by both analytic~\cite{cg,pps} 
and lattice methods~\cite{AliKhan:2001bx}, and the differences 
are not large enough to account for the discrepancy. 
This means that the difference between the lattice and dispersive 
values for ${\cal Q}_8$ represents a serious disagreement. It will 
be important to pursue an understanding of the physics behind this 
disagreement, as in principle both methods are fully rigorous.

In the case of the large $N_c$ evaluation~\cite{knecht}, there 
are differences in principle
with our approach because they have not performed the complete
matching at two loops. In addition, instead of using data for the
evaluation of the vacuum polarization functions, they use a vector meson
dominance model. Nevertheless, their results are consistent 
with ours, especially once the NLO radiative corrections 
are included in their work.

A recent manuscript~\cite{lund} also discusses
matching and dimension-eight effects, but in the context of an 
`X-boson method'. 
The greatest procedural difference with our work comes in the evaluation
of the ${\cal Q}_8$ matrix element. They find small non-factorizeable effects
such that their dominent contributions are the factorized terms, refined
to include the scale dependence from the matching procedure.

The VSA values are obtained from 
\beq
\langle (\pi\pi)_{I=2} | {\cal Q}_8 |K^0\rangle_{\rm VSA} = 
3 \langle (\pi\pi)_{I=2} | {\cal Q}_7 |K^0\rangle_{\rm VSA} = 
{2 F_\pi M_K^4 \over (m_s + m_d)_{\mu = 2~{\rm GeV}}^2 } 
\simeq 0.94~{\rm GeV}^3 \ \ , 
\label{vsa}
\eeq
where we take $(m_s + m_d)_{\mu = 2~{\rm GeV}} = 110~{\rm MeV}$ 
based on the value of $m_s$ given by the midpoint of the combined 
sum rule and lattice ranges in the recent review Ref.~\cite{gm} 
in combination with quark mass ratios obtained in ChPT 
analyses~\cite{hl}. 

The matrix element for ${\cal Q}_7$ has little phenomenological
interest, and will be useful primarily for the comparison with the
lattice as both the dispersive and lattice evaluations improve.
However the matrix element for ${\cal Q}_8$ is one of the main
contributions to $\epsilon'/\epsilon$, and gives a negative 
contribution to that quantity. Our rather large result implies 
that the electroweak penguin contribution is
\beq
{\epsilon' \over \epsilon}\bigg|_{\rm EWP} 
= \left( -2.2 \pm 0.7 \right) \times 10^{-3} \ \ .
\eeq
The gluonic penguin matrix element needs to be large enough to bring the
result up to the experimental value. Given this important consequence,
the resolution of the disagreement of the lattice and dispersive values
will be specially interesting.

\vfill\eject
\begin{table*}[t]
\caption{Comparison of Matrix Element Determinations at 
$\mu = 2$~GeV.\label{tab:compare}}
\vspace{0.4cm}
\begin{center}
\begin{tabular}{l|cc|cc}
& \multicolumn{2}{c}{$\langle (\pi\pi)_{I=2} | {\cal Q}_7 |
K^0\rangle({\rm GeV}^3)$} &
\multicolumn{2}{c}{$\langle (\pi\pi)_{I=2} | {\cal Q}_8 |
K^0\rangle({\rm GeV}^3)$} \\ \hline
Method of Calculation & NDR & HV
& NDR & HV \\ \hline
This work & $0.16 \pm 0.10$ & $0.49 \pm 0.07$
& $2.22 \pm 0.67$ & $2.46 \pm 0.70$ \\
Lattice~\cite{donini} & $0.11 \pm 0.04$ & $0.18 \pm 0.06$
& $0.51 \pm 0.10$ & $0.57 \pm 0.12$ \\
Large $N_c$~\cite{knecht} & $0.11 \pm 0.03$ & $0.67 \pm 0.20$
& $3.5 \pm 1.1$ & $3.5 \pm 1.1$ \\
X-boson~\cite{lund} & $0.26 \pm 0.03$ & $0.39 \pm 0.06$  
& $1.2 \pm 0.5$ & $1.3 \pm 0.6$ \\
VSA & $0.32$ & $0.32$  & $0.94$ & $0.94$ \\
\end{tabular}
\end{center}
\end{table*}

\acknowledgements
This work was supported in part by the National
Science Foundation under Grant PHY-9801875.
The work of V.C. is supported by TMR, EC-Contract No. ERBFMRX-CT980169
(EURODA$\Phi$NE).
KM would like to acknowledge the ongoing
support of the Natural Sciences and Engineering Research
Council of Canada, the CSSM at the University of Adelaide and the 
Theory Group at TRIUMF, as well as 
useful conversations with Andreas H\"ocker
and Shaomin Chen concerning the ALEPH data tabulations.

\eject

\end{document}